\documentclass{icrc2009}

\usepackage{graphicx}   
\usepackage{caption}    
\usepackage{subfig}
\usepackage{fixltx2e}
\usepackage{url}

\newcommand{\shorttitle}[1]%
{\markboth{Proceedings of the 31\MakeLowercase{$^{st}$} ICRC, {\L}\'{o}d\'{z} 2009}{#1} }
\newcommand{\etal}{\MakeLowercase{\textit{et al. }}} 

\begin{document}
\title{Observation of the Galactic Cosmic Ray Moon shadowing effect with the ARGO-YBJ experiment}

\author{\IEEEauthorblockN{Roberto Iuppa\IEEEauthorrefmark{1}
                           \IEEEauthorrefmark{2},
                           Daniele Martello\IEEEauthorrefmark{3}\IEEEauthorrefmark{4},
                           Bo Wang\IEEEauthorrefmark{5} and
                           Giovanni Zizzi\IEEEauthorrefmark{3}\IEEEauthorrefmark{4}\\
                           on behalf of the ARGO-YBJ Collaboration
}
                            \\
\IEEEauthorblockA{\IEEEauthorrefmark{1} INFN Sezione Roma Tor
Vergata, via della Ricerca Scientifica 1, Rome - Italy}
 \IEEEauthorblockA{\IEEEauthorrefmark{2} Dipartimento di Fisica,
 Universit\'a Roma Tor Vergata, via della Ricerca Scientifica 1, Rome - Italy}
 \IEEEauthorblockA{\IEEEauthorrefmark{3} INFN Sezione di Lecce, via
per Arnesano, 73100 Lecce - Italy}
 \IEEEauthorblockA{\IEEEauthorrefmark{4} Dipartimento di Fisica
dell'Universit\`a del Salento, via per Arnesano, 73100 Lecce -
Italy}
 \IEEEauthorblockA{\IEEEauthorrefmark{5} Key Laboratory of Particle
Astrophysics, IHEP - Chinese Academy of Science, P.O. Box 918,
100049 Beijing, P.R. China}}

\shorttitle{R. Iuppa \etal Moon Shadow Observation}
\maketitle

\begin{abstract}

Cosmic rays are hampered by the Moon and a deficit in its
direction is expected (the so-called \emph{Moon shadow}). The Moon
shadow is an important method to determine the performance of an
air shower array. In fact, the westward displacement of the shadow
center, due to the propagation of cosmic rays in the geomagnetic
field, allows to calibrate the energy scale of the primary
particles observed by the detector. In addition, the shape of the
shadow allows a measurement of the angular resolution and the
position of the deficit at high energy allows the evaluation of
the pointing accuracy of the detector.

In this paper we present the observation of the galactic cosmic
rays Moon shadowing effect performed by the ARGO-YBJ experiment in
the multi-TeV energy region. The
measured angular resolution as a function of the shower size is
 compared with the expectations from a MC simulation.
  \end{abstract}


\section{Introduction}

The angular resolution is a critical feature of an Extensive Air
Shower (EAS) array in gamma-ray astronomy. In fact, the rejection
of the nearly isotropic background of charged cosmic rays is
mainly performed by improving the angular resolution, thus
reducing the source region extension. Hence the tuning of a firm
calibration technique of the angular resolution is mandatory.

The CYGNUS experiment in 1991 reported the first determination of
the angular resolution of an EAS detector by exploiting the
analysis of the shadow of the Moon \cite{cygnus}. In fact, as
pointed out in 1957 by Clark \cite{clark}, the cosmic rays are
hampered in their propagation to the Earth due to the Moon's
presence and a deficit of events in its direction is expected: the
so-called \emph{Moon shadow}.

At high energy, the Moon shadow would be observed by an ideal
detector as a 0.26$^{\circ}$ wide circular deficit of events,
centered on the Moon position. The deviation from such an ideal
case gives us information about the Point Spread Function (PSF) of
the detector. The shape of the deficit allows the measurement of
the angular resolution and its position allows the evaluation of
the absolute pointing accuracy of the detector. In addition,
positively charged particles are 
eastward deflected, due to the geomagnetic field, with an energy dependence 
$\Delta\theta\sim 1.6^{\circ}Z/E_{TeV}$. As a consequence, the
observation of the displacement of the Moon provides a direct
check of the relation between shower size and primary energy.
Therefore, the analysis of the Moon shadow allows the calibration
of the performance of an EAS array.

The same shadowing effect can be observed in the direction of the
Sun but the interpretation of the shadowing phenomenology is more
complex. In fact, the displacement of the shadow from the apparent
position of the Sun could be explained by the joint effects of the
geomagnetic field and of the solar and Interplanetary Magnetic
Fields, whose configuration considerably changes with the phases
of the solar activity cycle \cite{amenomori-sun00}. Results about
the Sun shadow observation with the ARGO-YBJ experiment are
discussed in \cite{argo-sun09}.

In this paper we present the observation of the galactic cosmic
rays Moon shadowing effect carried out by the ARGO-YBJ experiment.
We report on the angular resolution of the detector in the multi-TeV 
energy region. The
pointing error is also investigated. 
\section{The ARGO-YBJ experiment}

The ARGO-YBJ detector, located at the YangBaJing Cosmic Ray
Laboratory (Tibet, P.R. China, 4300 m a.s.l.), is the only
experiment exploiting the \emph{full coverage} approach at very
high altitude. The detector is constituted by a central carpet
$\sim$74$\times$78 m$^2$, made of a single layer of Resistive
Plate Chambers (RPCs) with $\sim$92$\%$ of active area, enclosed
by a partially instrumented guard ring that extends the detector
surface up to $\sim$100$\times$110 m$^2$. The apparatus has a
modular structure, the basic data acquisition element being a
cluster (5.72$\times$7.64 m$^2$), namely a group of 12 RPCs
(2.80$\times$1.25 m$^2$ each). Each chamber is read by 80 strips
of 7$\times$62 cm$^2$ (the spatial pixel), logically organized in
10 independent pads of 56$\times$62 cm$^2$ representing the time
pixel of the detector. The RPCs are operated in streamer mode with
a standard gas mixture (Argon 15\%, Isobutane 10\%,
TetraFluoroEthane 75\%), the High Voltage settled at 7.2 kV
ensures an overall efficiency of about 96\% \cite{aielli06}. The
central carpet contains 130 clusters (hereafter ARGO-130) and the
full detector is composed of 153 clusters for a total active
surface of $\sim$6700 m$^2$.

All events giving a number of fired pads N$_{pad}\ge$ N$_{trig}$
in the central carpet within a time window of 420 ns are recorded.
The spatial coordinates and the time of any fired pad are then
used to reconstruct the position of the shower core and the
arrival direction of the primary.

The ARGO-YBJ experiment started recording data with the whole
central carpet in June 2006. Since 2007 November the full
detector is in stable data taking at the multiplicity trigger
threshold N$_{trig}\geq$20 and a duty cycle $\sim 90\%$: the
trigger rate is about 3.6 kHz.

The reconstruction of shower parameters is split into the
following steps. First the shower core position is derived with
the Maximum Likelihood method from the lateral density
distribution of the secondary particles. In the second step, given
the shower core position, the shower axis is reconstructed by
means of an iterative weighted planar fit being able to reject
the time values belonging to the non-gaussian tails of the arrival
time distributions. A conical correction with a slope fixed to
$\alpha$ = 0.03 rad is applied to the surviving hits in order to
improve the angular resolution \cite{argo-rec}.

\section{Monte Carlo Simulation}

The air showers development in the atmosphere has been generated
with the CORSIKA v. 6.500 code including the QGSJET-II.03 hadronic
interaction model for primary energy above 80 GeV and the FLUKA
code for lower energies \cite{corsika}. Cosmic ray spectra of p,
He and CNO have been simulated in the energy range from 30 GeV to
1 PeV following \cite{wiebel-sooth}. The relative fractions (in \%
of the total) after triggering by the ARGO-YBJ detector for events
with N$_{strip}\geq$30 are: p$\sim$88\%, He$\sim$10\%,
CNO$\sim$2\%. About 3$\cdot$10$^{11}$ showers have been
distributed in the zenith angle interval 0-60 degrees. The
secondary particles have been propagated down to a cut-off energy
of 1 MeV. The experimental conditions have been reproduced via a
GEANT3-based code. The shower core positions have been randomly
distributed sampling in energy-dependent area up to
10$^3\times$10$^3$ m$^2$, centered on the detector.

\section{Data analysis}

For the analysis of the shadowing effect a 10$^{\circ}\times$
10$^{\circ}$ sky map in celestial coordinates (right ascension and
declination) with 0.1$^{\circ}\times$0.1$^{\circ}$ bin size,
centered on the Moon location, is filled with the detected events.
The background is evaluated with both the \emph{time swapping}
\cite{alexandreas} and the \emph{equi-zenith angle}
\cite{amenomori93} methods.

With the time swapping method, N \emph{"fake"} events are
generated for each detected one, by replacing the measured arrival
time with new ones. These events are randomly selected within a 3
hours wide buffer of recorded data. Swapping the time is swapping
the right ascension, keeping unchanged the declination. A new sky
map (background map) is built by using 10 such fake events for
each real one, so that the statistical error on the background can
be kept small enough.

With the equi-zenith angle method 6 off-source bins are
symmetrically aligned on both sides of the on-source field, at the
same zenith angle. The off-source bins are each set at an azimuth
distance 5$^{\circ}$/sin$\theta$ from the on-source bin, where
$\theta$ is the zenith angle of the Moon position. Other
off-source bins are located every 5$^{\circ}$/sin$\theta$ from the
nearest off-source bins. The average of the event densities inside
these bins was taken to be the background.

To maximize the signal to noise ratio, the bins are then grouped
over a circular area of radius $\psi$, i.e. every bin is filled
with the content of all the surrounding bins whose center is
closer than $\psi$ from its center. The value of $\psi$ is related
to the angular resolution of the detector, and corresponds to the
radius of the observational window that maximizes the signal to
noise ratio, which in turn depends on the number of fired pads of
the event: when the PSF is a Gaussian with rms $\sigma$, $\psi=
\sigma\cdot 1.58$ and contains $\sim$72$\%$ of the events.
Finally, the integrated background map is subtracted from the
corresponding integrated event map, thus obtaining the "source
map". For each bin of such a map, the deficit significance with
respect to the background is calculated according to the Li and Ma
formula \cite{li-ma}. Notice that in the integrated maps
neighboring bins are correlated.

 \begin{figure}[!t]
  \centering
  \includegraphics[width=3.0in]{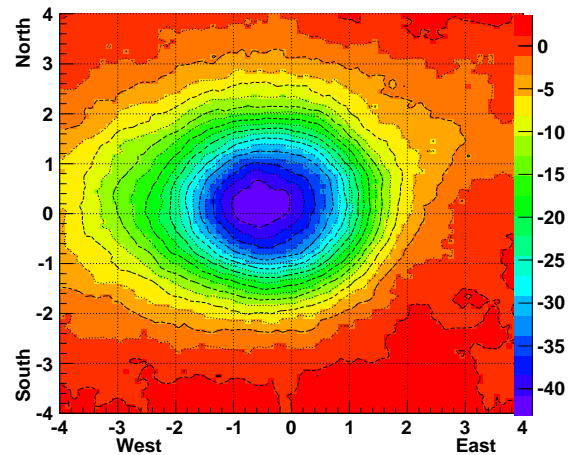}
  \caption{Moon shadow significance map observed by the ARGO-YBJ
detector in 2063 hours on-source for events with N$_{strip}\geq$30
and zenith angle $\theta<50^{\circ}$. The color scale gives the
statistical significance.}
  \label{fig:moon-wangbo}
 \end{figure}

The analysis reported in this paper refers to events collected
after the following event selection: (1) each event should fire
more than 30 strips on the ARGO-130 central carpet to avoid any
threshold effect; (2) the zenith angle of the shower arrival
direction should be less than 50$^{\circ}$; (3) the reconstructed
core position should be inside an area 250$\times$250 m$^2$
centered on the detector; (4) the reduced $\chi^2$ of the final
temporal fit should be less than 100 ns$^2$. According to our
simulation studies, the median energy of the selected protons
firing 30 $\div$ 60 strips is E$_{50}\approx$1.4 TeV (mode energy
$\sim$0.30 TeV).

In Fig.\ref{fig:moon-wangbo} the Moon shadow observed with all
data recorded since June 2006 (2063 hours on-source) for events
with N$_{strip}\geq$30 and zenith angle $\theta<50^{\circ}$ is
shown. The statistical significance of the observation is about 43
standard deviations.

\section{Results}

\begin{figure}[!t]
  \centering
  \includegraphics[width=3.0in]{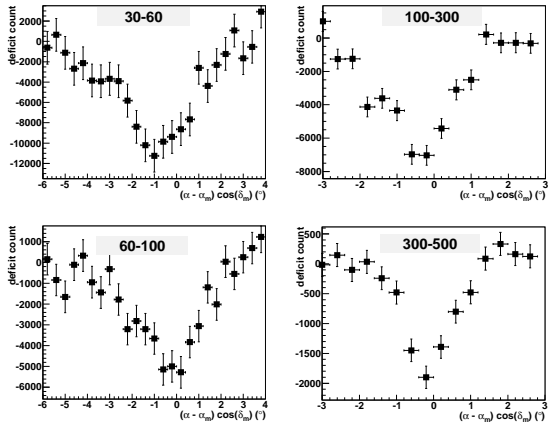}
\caption{Deficit counts measured around the Moon projected along
the East-West axis for different multiplicity bins.}
  \label{fig:proj-ew}
 \end{figure}

The deficit counts observed around the Moon projected to the
East-West axis are shown in Fig. \ref{fig:proj-ew} for 4
multiplicity bins. We use the events contained in an angular slice
parallel to the East-West axis and centered to the observed Moon
position. The widths of these bands are function of the
N$_{strip}$-dependent angular resolution: $\pm$3.3$^{\circ}$ in
30$\leq N_{strip} <$ 60, $\pm$2.6$^{\circ}$ in 60$\leq N_{strip}
<$ 100, $\pm$2.0$^{\circ}$ in 100$\leq N_{strip} <$ 300,
$\pm$1.5$^{\circ}$ in 300$\leq N_{strip} <$ 500. As an expected
effect of the geomagnetic effect, the profile of the shadow is
broadened and the peak positions shifted westward as the
multiplicity (i.e., the cosmic ray primary energy) decreases. We note
that in the lowest multiplicity bin (N$_{strip}$ = 30 - 60) the
Moon is shifted by about 1$^{\circ}$, as expected for a primary 
of rigidity $1.6\textrm{TeV}/Z$: this is the first time that
an EAS-array is able to detect showers with such a low primary
energy.

 \begin{figure}[!t]
  \centering
  \includegraphics[width=3.0in]{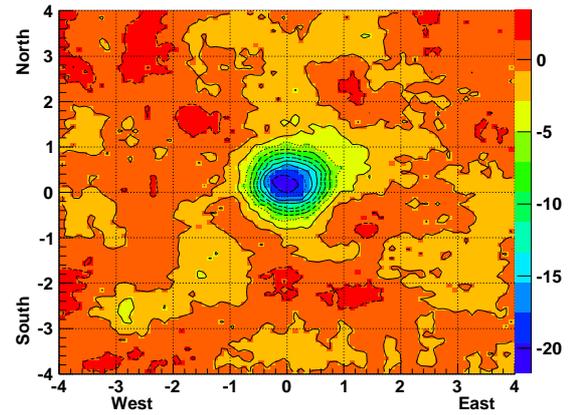}
  \caption{Moon shadow significance map observed by the ARGO-YBJ
detector in 2063 hours on-source for events with
N$_{strip}\geq$1000 and zenith angle $\theta<50^{\circ}$. The
color scale gives the statistical significance.}
  \label{fig:moon-hemap}
 \end{figure}

The best procedure to evaluate the pointing accuracy is 
to observe the position of the Moon shadow produced by 
high-energy cosmic rays which are negligibly affected by the
geomagnetic field.
For protons of 30 TeV we expect a deflection of about
0.05$^{\circ}$. For heavier nuclei this deflection will increase
but as the composition of cosmic rays in this energy range is
dominated by the light component (nuclei heavier than CNO
contribute to the rate less than 3\% in the whole strip
multiplicity range \cite{catalano}) we expect only a small
contribution from heavy ions to the blurring of the Moon shadow.
As can be seen from the Fig. \ref{fig:moon-hemap}, the observed
high energy (N$_{pad}\geq$1000, E$^{50}_p\sim$ 30 TeV) Moon shadow
position is centered in the East-West direction but we observe a
residual shift towards the North. Since the displacement along the
North-South axis is not affected by the geomagnetic field at the
Yangbajing latitude \cite{mcmoon}, we are able to investigate this
pointing error without the Moon shadow simulation as a function of
the multiplicity.
The analysis has been performed both with the time-swapping and
the equi-zenith angle methods: the results are in good agreement
and suggest that there is a residual systematic shift towards
North of (0.20$\pm$0.05)$^{\circ}$, independent of multiplicity.
As a conservative estimate we assume our systematic errors to coincide 
with this displacement in both North-.South and East-West projection. 
These upper limits for the systematic errors are however much
smaller than our angular resolution, at least for the bulk of
data, and can therefore be neglected in the point source searches.

\begin{figure}[!t]
  \centering
  \includegraphics[width=3.0in]{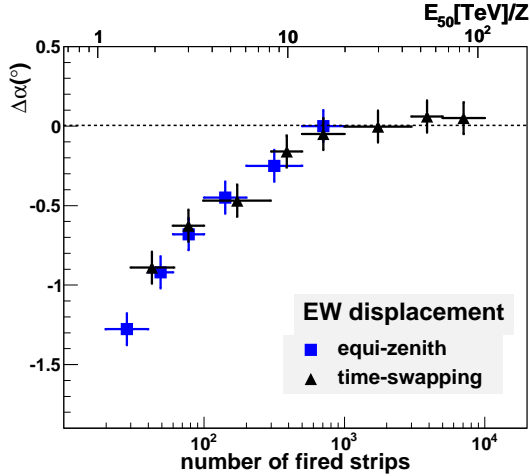}
\caption{Observed displacement of the Moon shadow in the East-West
direction as a function of multiplicity. The upper scale refers to
the median energy of rigidity (TeV/Z) in each multiplicity bin
(shown by the horizontal errors). The experimental data calculated
with two different methods (see text).}
 \label{fig:moonshift}
 \end{figure}

In the Fig. \ref{fig:moonshift} the displacement of the Moon
shadow in the East-West direction is shown. Two different methods agree with 
each other quite well. A comparison between the measurement and the 
simulations allows to attribute this displacement to an absolute energy calibration 
of the detector.
\begin{figure}[!t]
  \centering
  \includegraphics[width=3.0in]{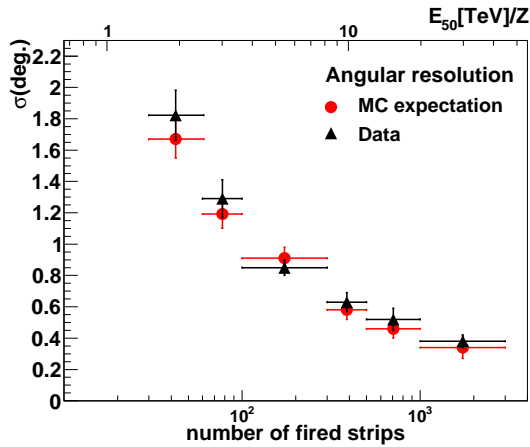}
  \caption{Measured angular resolution of the ARGO-YBJ detector
compared to expectations from MC simulation as a function of the
multiplicity. The upper scale refers to the median energy of
rigidity (TeV/Z) in each multiplicity bin (shown by the horizontal
errors).} \label{fig:angresol}
 \end{figure}

The PSF of the detector, studied in the North-South projection, not
affected by the geomagnetic field, is Gaussian for
N$_{strip}\ge$100, while for lower multiplicities it can be
described with an additional Gaussian, which contributes for about
20\%. For these events the angular resolution is calculated as the
weighted sum of the $\sigma^2$ of each gaussian.
In Fig. \ref{fig:angresol} the measured angular resolution is
compared to expectations from a MC simulation as a function of the
multiplicity. As can be seen, the values are in fair agreement:
the angular resolution $\sigma$ of the ARGO-YBJ experiment for
cosmic ray-induced air showers is less than 0.6$^{\circ}$ for
N$_{strip}\geq$300. The effect of the finite angular width of the
Moon on the angular resolution (less than 5\% if
$\sigma>$0.4$^{\circ}$ and only 1.7\% if $\sigma>$0.7$^{\circ}$)
is discussed in \cite{mcmoon}. From MC simulations we draw that
the angular resolution for $\gamma$-induced showers at the
threshold is at least 30\% lower due to their better defined time
profile \cite{argo-rec}. A new reconstruction algorithm based on a
weighted fit of the time profile in under study in order to
improve the angular resolution.

A new reconstruction algorithm based on a weighted fit of the time 
profile is under study in order to improve the angular resolution.

\section{Conclusions}

The galactic cosmic ray Moon shadowing effect has been observed by
the ARGO-YBJ experiment in the multi-TeV energy region with high
statistical significance. The analysis has been performed both
with the time-swapping method and the equi-zenith angle method in
order to investigate possible biases in the background
calculation. The measured angular resolution is in good agreement 
with MC simulations, making us confident in the reconstruction 
algorithms, 
we can further find an absolute energy calibration of the detector.

\end{document}